\documentclass{PoS}

\title{Single-nucleon experiments}

\ShortTitle{Single-nucleon experiments}

\author{\speaker{Alexandre Deur}%
         \thanks{Acknowledgments: This work is supported by the U.S. 
Department of Energy (DOE). The Jefferson Science Associates (JSA) 
operates the Thomas Jefferson National Accelerator Facility for the DOE 
under contract DE-AC05-84ER40150. }\\
        Jefferson Lab\\
        E-mail: \email{deurpam@jlab.org}}

\abstract{We discuss the Jefferson Lab low momentum transfer data on moments of the nucleon spin structure functions $g_1$ and $g_2$ and on single charged pion electroproduction off polarized proton and polarized neutron. 
A wealth of data is now available, while more is being analyzed or expected to be taken in the upcoming years. Given the low momentum transfer  selected by the experiments, these data can be compared to calculations from Chiral Perturbation theory, the effective theory of strong force that should describe it at low momentum transfer. 
The data on various moments and the respective calculations do not consistently agree. In particular, experimental data for higher moments disagree with the calculations.The absence of contribution from the $\Delta$ resonance 
in the various observables was expected to facilitate the calculations and hence make the theory predictions 
either more robust or valid over a larger $Q^2$ range.
Such expectation is verified only for the Bjorken sum, but not for other observables in which the $\Delta$  is suppressed.  Preliminary results on pion electroproduction off polarized  nucleons are also presented and compared to phenomenological models for which contributions from different resonances are varied. Chiral Perturbation calculations of these observables, while not yet available, would be valuable and, together with these data, would provide an extensive test of the effective theory.}

\FullConference{6th International Workshop on Chiral Dynamics\\
                 July 6-10 2009\\
                 Bern, Switzerland}

\begin{document}

\section{Introduction and context}

Understanding Nature using the technique of particle scattering is a vast endeavor. As the 
title of this contribution indicates, we will focus on scattering off the nucleon. This is still a 
large subject. Consequently, in addition to \emph{single} 
nucleon, we will mostly restrict this presentation to  \emph{single} particle detection (i.e. inclusive 
scattering) from a \emph{single} type of beam (electron scattering), with \emph{single} spin 
directions for the beam and target (i.e doubly polarized scattering) and we will mostly report on 
results from a \emph{single} laboratory (Jefferson Lab).  

Doubly polarized electron scattering off the nucleon is a powerful tool to understand strong 
interactions: The polarization provides the most stringent constraints on the theory, while
the leptonic probe is the cleanest way to access the structure of the nucleon. 
Having a single nucleon target removes the difficulties arising from collective effects, such as the EMC 
effects~\cite{EMC}. Since the nucleon structure is ruled by strong interactions, an understanding of its
structure translates to understanding of strong interactions. The gauge theory of  strong
interactions is Quantum Chromodynamics (QCD). Compared to other interactions, it stands out because it is generally 
non-perturbative, except for high energy ($\gg$ GeV) reactions. Consequently, the first natural step toward 
our understanding of strong interactions is to check that QCD is valid where we can analytically solve it, while
effective theories (e.g. Chiral Perturbation Theory) or numerical methods (Lattice QCD) need to be developed 
and tested to cover the lower energy region where QCD is non-perturbative.
For final completion of our understanding, one then needs to connect the fundamental and effective
theories, just like the empirical rules of chemistry  have been fundamentally justified by quantum (atomic) physics, or geometrical 
optics by electromagnetism. Part of the first step -the validity of perturbative QCD (pQCD) even when spin degrees
of freedom are explicit- has been 
achieved by a first generation of experiments that ran in the 1980's-1990's at facilities such as SLAC, CERN or DESY. Here, 
we will discuss the experimental program at Jefferson Lab to achieve the other part of the first step 
-testing effective theories and numerical methods-, and the second step: the connection 
between pQCD and the effective descriptions.

\section{Jefferson Lab, the experimental halls and their polarized targets}

Jefferson Lab is an accelerator facility delivering a high quality continuous electron beam with 
energy up to 6 GeV. Beam polarization reaches 85\% and beam current can be up to 200 $\mu$A. Experiments are carried out in three experimental halls: Halls A, B and C. Hall A~\cite{Hall A} and C contain high resolution spectrometers for high precision experiments with limited phase space coverage. Hall B contains a large acceptance spectrometer~\cite{Hall B} for exclusive experiments and/or exploratory measurements over a wide
kinematic range.
Each hall is home of polarized targets. Hall A is home of a $^3$He gaseous target polarized by
optical pumping. Polarized $^3$He acts as an effective polarized neutron target because the dominant nuclear state is the S state, for which the Pauli principle forces the two proton spins to be anti-aligned. Hence, the single neutron contributes dominantly (about 90\% ) to the target polarization.
The Hall A target, together with the JLab continuous beam, achieves the largest polarized luminosity in the world ($10^{36}$ s$^{-1}$cm$^{-2}$). This,
with its low dilution from unpolarized materials (typically, about 30\% of the events comes from polarized neutron) and its excellent polarization (60-70\%) allows for high precision
experiments. The target polarization can be oriented in any direction, including the vertical one. In this talk, we
focus on neutron results although data on $^3$He structure are available as well. This topic is
covered by K. Slifer's contribution to this conference. Hall B and Hall C employ solid targets polarized by DNP 
(Dynamical Nuclear Polarization). The materials most often used are ammonia $^{15}$NH$_3$ and $^{15}$ND$_3$.
 Hence the proton, deuteron and neutron structures are studied in halls B and C. The targets allow to reach
 good luminosities  ($10^{34}$ s$^{-1}$cm$^{-2}$ for Hall B and $10^{35}$ s$^{-1}$cm$^{-2}$ for Hall C). The somewhat
 lower luminosity in Hall B is compensated by the large acceptance of the CLAS detector. Solid ammonia targets
 have relatively high dilution from unpolarized materials (e.g. typically, about 15\% of the events comes from polarized proton) but reach high polarizations (90\% for NH$_3$ and 40\%
 for ND$_3$). The Hall B target~\cite{Hall B DNP target} is polarized along the beam direction, while the Hall C target polarization can
 be along the beam direction or transverse (in the horizontal plane) to it. Hall B also is also home of the FROST~\cite{FROST} polarized target, but we will not discuss
 it since it cannot accommodate electron beams. The former LEGS polarized HD target~\cite{HDice} will also be available soon in Hall B.
 It is presently being redesigned in order to accommodate electron beams in addition to its original usage with
 photon beams. If this target does stand electron beams, an important  possibility of a transversely polarized target (with low dilution from unpolarized materials) experiments in Hall B will be opened.
 
 \section{Sum rules and the spin structure of the nucleon}
The information on the longitudinal spin structure of the nucleon is contained in the
$g_1(x,Q^2)$ and $g_2(x,Q^2)$ spin structure functions. The kinematic variable  
$Q^2$ is the squared four-momentum transfered from the beam to the target. It fixes
the space-time scale at which the nucleon is probed. The other kinematics variable,  
$x=Q^2/(2M \nu)$, is the Bjorken scaling variable ($\nu$ is the energy transfer from the beam to 
the target, and $M$ is the nucleon mass). The variable $x$  is interpreted in the parton model as the fraction of nucleon momentum carried by 
the struck quark. Another kinematics variable that we will use in this talk is $W$ the invariant mass of
the recoiling system: $W^2=M^2+2M\nu-Q^2=M^2+Q^2(1/x-1)$.

Although  spin structure functions are the basic observables for nucleon longitudinal spin studies, 
we will focus on their integrals formed over $x$ and weighted by powers of $x$. Considering these moments 
is advantageous because of the resulting simplifications. More importantly, such integrals are at the core
of the dispersion relation formalism. Dispersion relations  relate the integral over
the imaginary part of a quantity to its real part. Expressing the 
imaginary part as a function of the real part using the optical theorem yields 
\emph{sum rules}. When additional hypotheses are used, such as a low energy theorem,
or the validity of Operator Product Expansion (OPE), the sum rules then relate the integral 
to a static property of the target, e.g. its anomalous magnetic moment, an electromagnetic
polarizability, or its axial charge. If the static property is well known (e.g. the anomalous moment or
the axial charge), the verification of the sum rule provides a check of the theory and hypotheses used in the 
sum rule derivation. When the static property is not known because for instance it is difficult to 
measure directly (e.g. the generalized electromagnetic polarizabilities), sum rules can be used to access 
them. In that case, the theoretical framework used to derive the sum rule has to be assumed to be valid. Details on integrals of spin structure functions and sum rules are given e.g. in the review~\cite{review moments}. 

Several spin sum rules exists. We
will focus on the Bjorken sum rule~\cite{Bjorken}, the Gerasimov-Drell-Hearn (GDH) sum rule~\cite{GDH} and spin polarizability sum rules.  In this paper, we will consider the $n$-th Cornwall-Norton moments: $\int_{0}^{1}dx g_{1}^{N}(x,Q^2) x^n$, with
$N$ standing for proton or neutron. We write the first moments as $\Gamma_{1}^{N}(Q^2)\equiv\int_{0}^{1}dx g_{1}^{N}(x,Q^2)$.

\section{The generalized Bjorken and GDH sum rules}
The Bjorken sum rule~\cite{Bjorken} relates the integral over the isovector part of the first
spin structure function, $\int_{0}^{1}dx(g_1^p-g_1^n)$, 
to the nucleon axial charge $g_A$. The original sum rule stands at infinite $Q^2$ but has been generalized
to finite $Q^2$ with the OPE (i.e. pQCD).   This relation has been essential for understanding the 
nucleon spin structure and establishing, \emph{via} its 
$Q^2$-dependence, that QCD describes 
the strong force even when spin degrees of freedom are explicit.
The Bjorken integral has been measured in polarized deep inelastic
lepton scattering (DIS) at SLAC, CERN and DESY~\cite{e142}-\cite{HERMES}
and at moderate and low $Q^2$ at Jefferson Lab (JLab)~\cite{eg1a proton}-\cite{RSS}. A recent review of these
data can be found in~\cite{review KCL}.
The OPE yields the following  expression for the sum rule:
\begin{eqnarray}
\label{eq:bj(Q2)}
\Gamma_{1}^{p-n}(Q^2)\equiv\int_{0}^{1}dx
\left( g_{1}^{p}(x,Q^2)-g_{1}^{n}(x,Q^2) \right)=
\hspace{1cm}\\
\frac{g_{A}}{6}\left[1-\frac{\alpha_{s}}{\pi}-3.58
\frac{\alpha_{s}^{2}}{\pi^{2}}-
20.21\frac{\alpha_{s}^{3}}{\pi^{3}}+...\right]+
\sum_{i=2}^{\infty}{\frac{\mu_{2i}^{p-n}(Q^{2})}{Q^{2i-2}}} \nonumber
\end{eqnarray}
where  $\alpha_s(Q^2)$ is the strong coupling constant. The bracket term (known as the leading twist term) is mildly dependent on $Q^2$ due to soft gluon radiations. The summation term contains non-perturbative power corrections (higher twists). These are quark and gluon correlations describing the nucleon structure away from the large $Q^2$ (small distances) limit.

The generalized Bjorken sum rule has been derived for small distances. For large distances, at $Q^2=0$, one finds the GDH sum rule~\cite{GDH}. For a spin 1/2 target, it reads:
\begin{equation}
\int_{\nu_{0}}^{\infty}d\nu\frac{\sigma_{1/2}(\nu)-\sigma_{3/2}(\nu)}
{\nu}=-\frac{2\pi^{2}\alpha\kappa^{2}}{M_t^{2}}\label{eq:gdh}
\end{equation}
where $\nu_{0}$ is the pion photoproduction threshold, $\sigma_{1/2}$ and 
$\sigma_{3/2}$ are the helicity dependent photoproduction cross sections when
the sum of the photon and target helicities is 1/2 and 3/2, respectively. $\kappa$ is the anomalous
magnetic moment of the target and $M_t$ its mass. $\alpha$
is the fine structure constant. The GDH sum rule can also be written for target of any spin (e.g. a deuteron target).  
Replacing the photoproduction cross sections by the electroproduction ones 
generalizes the left hand side of Eq.~\ref{eq:gdh} to any $Q^2$, that is, it generalizes the sum. Such generalization 
depends on the choice of convention for the virtual photon flux, see e.g. 
ref.~\cite{review moments}. X. Ji and J. Osborne~\cite{ji01} showed that the 
sum \emph{rule} itself  (i.e. the whole Eq.~\ref{eq:gdh}) can be generalized as:
\begin{equation}
\frac{8}{Q^2}\int_0^{1^-} dx~g_1=S_1(0,Q^2)\label{eq:gdh*}
\end{equation} 
where $S_1(\nu,Q^2)$ is the spin-dependent Compton amplitude and the $1^- \equiv 1- \epsilon$ integration upper limit excludes the elastic contribution. This generalization
of the GDH sum rule makes the connection between the Bjorken and GDH generalized sums evident: 
\begin{equation}
GDH=\frac{Q^2}{8} \times Bjorken, \label{eq:gdh-bj}
\end{equation} 
where the elastic contribution to the Bjorken sum, Eq.~\ref{eq:bj(Q2)}, is in this case excluded. The connection between the GDH and Bjorken sum rules makes theories available  to compute the moment $\Gamma_1$ at any $Q^2$. The Bjorken sum rule, evolved with $Q^2$ according to pQCD provides the theoretical prediction at 
large $Q^2$: Calculation of the  spin dependent Compton amplitude $S_1(\nu,Q^2)$ with
Lattice QCD (intermediate $Q^2$) or at low $Q^2$ with Chiral Perturbation Theory ($\chi$pT, the effective 
theory of strong force at large distances) yields the theoretical predictions at intermediate and low $Q^2$.
Thus, we are provided with a convenient observable to understand how the strong force transitions from its description in term of fundamental degrees of freedom (quarks and gluons; small distances) to its description
in term of effective degrees of freedom (hadrons; large distances).

\section{Experimental measurements of the first moments}

\begin{figure}
\includegraphics[width=1.05\textwidth]{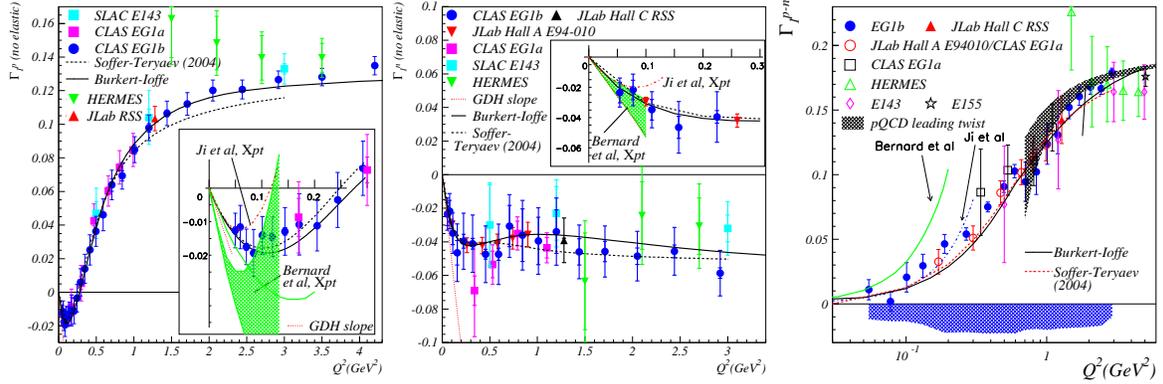}
\caption{(Color online) Experimental data from SLAC, CERN, DESY and JLab at low and intermediate $Q^2$ on $\Gamma_1^p$ (left), $\Gamma_1^n$ (center) and $\Gamma_1^{p-n}$ (right).}
\label{gammas}
\end{figure}

Results from experimental measurements from SLAC~\cite{SLAC}, CERN~\cite{SMC}, 
DESY~\cite{HERMES} and JLab~\cite{eg1a proton}-\cite{RSS} of the first moments $\Gamma_1$ are shown in 
Figure~\ref{gammas}.
There is an thorough mapping of the moments at intermediate $Q^2$ and enough data points
at low $Q^2$ to start testing $\chi PT$. In this context, the Bjorken sum is especially
important because the (p-n) subtraction largely cancels the $\Delta$ 
resonance contribution which should make the $\chi PT$ calculations significantly more 
reliable~\cite{Burkert Delta}.
The comparison between the data at low $Q^2$ and $\chi PT$ 
calculations  can be seen  on Fig.~\ref{gammas}. The calculations are done
to next-to-leading order, suing either an explicitly covariant formalism~\cite{meissner} or the Heavy Baryon approximation~\cite{Ji chipt}. The $\chi PT$ 
calculations do not compute the  slope of $\Gamma_1$ at $Q^2$=0, but takes it from the GDH sum rule 
prediction since it provides the derivative of  $\Gamma_1$ at $Q^2$=0 (see Eq.~\ref {eq:gdh-bj}).
Consequently, $\chi PT$ calculates the deviation from the slope and this is what one 
should test. A meaningful comparison is then provided by fitting the lowest $Q^2$ data points
using the form $\Gamma_1^N=\frac{\kappa_N^2}{8M^2}Q^2+aQ^4+bQ^6...$ and compare the
obtained value of $a$ to the values calculated from $\chi PT$. Such comparison has been 
carried out for the proton, deuteron~\cite{EG1b moments} and the Bjorken 
sum~\cite{Bj EG1b} (numerical values are given on Fig.~\ref{table}). These fits point out 
the importance of including a $Q^6$ 
term for $Q^2<0.1$ GeV$^2$. The $\chi PT$ calculations agree well with the measurements
on the individual nucleons up to $Q^2 \sim 0.08$ GeV$^2$ for the Ji. \emph{et al} calculations.  They agree with the measurement of the Bjorken sum over a larger $Q^2$ span: up to  $Q^2 \sim 0.3$ GeV$^2$ for the Ji. \emph{et al} calculations, in accordance with the discussion in~\cite{Burkert Delta}. Phenomenological models~\cite{AO},\cite{soffer} are in good agreement with the data over the whole $Q^2$ range.

\section{Spin polarizability sum rules}
Higher moments of $g_1$ and $g_2$ are connected by sum rules to electromagnetic polarizabilities.
Those characterize  the coherent response of the nucleon to photon absorption. They are defined using 
low-energy theorems in the form of a series expansion in the photon energy.  The first term of the 
series comes from the spatial distribution of charge and current (form 
factors) while the second term results from the deformation of these 
distributions induced by the photon (polarizabilities). Hence, 
polarizabilities are as important as form factors in understanding 
coherent nucleon structure. \emph{Generalized} spin polarizabilities 
describe the response to \emph{virtual} photons. Using a low energy theorem, the generalized forward spin polarizability $\gamma_{0}$ is defined as:
\begin{eqnarray}
\Re e[g_{TT}(\nu,Q^{2})-g_{TT}^{p\hat{o}le}(\nu,Q^{2})]= \label{eq:sr2} 
(\frac{2\alpha}{M^{2}})I_{TT}(Q^{2})\nu+\gamma_{o}(Q^{2})\nu^{3}+O(\nu^{5}),
\end{eqnarray}
where $g_{TT}$ is the spin-flip doubly-virtual Compton scattering 
amplitude, and $I_{TT}$ is the coefficient of the $O(\nu)$ term of the 
Compton amplitude which can be used to generalize the 
GDH sum rule to non-zero $Q^2$. We have $I_{TT}(Q^{2}=0)=\kappa /4$. In 
practice $\gamma_{0}$ can be obtained from a sum rule which has a 
derivation akin to that of the GDH sum rule:
 \begin{eqnarray}
\gamma_{0}(Q^{2}) & = & (\frac{1}{2\pi^{2}})\int_{\nu_{0}}^{\infty}\frac{\kappa(\nu,Q^{2})}{\nu}\frac{\sigma_{TT}(\nu,Q^{2})}{\nu^{3}}d\nu
\end{eqnarray}
where $2\sigma_{TT}\equiv\sigma_{1/2}-\sigma_{3/2}$. We can express the sum rule in terms of the spin structure functions as:
\begin{eqnarray}
\gamma_{0}(Q^{2})=\frac{16\alpha M^{2}}{Q^{6}}\int_{0}^{x_{0}}dx~x^{2}\left(g_{1}-
\frac{4M^{2}}{Q^{2}}x^{2}g_{2}\right).
\label{eq:srg0}
\end{eqnarray}
Similar relations define the generalized longitudinal-transverse polarizability 
$\delta_{LT}$ :
\begin{eqnarray}
\Re e[g_{LT}(\nu,Q^{2})-g_{LT}^{p\hat{o}le}(\nu,Q^{2})]=
(\frac{2\alpha}{M^{2}})QI_{LT}(Q^{2})+Q\delta_{LT}(Q^{2})\nu^{2}+O(\nu^{4}),
\end{eqnarray}
\begin{eqnarray}
\delta_{LT}(Q^{2}) & = & (\frac{1}{2\pi^{2}})\int_{\nu_{0}}^{\infty}\frac{\kappa(\nu,Q^{2})}{\nu}\frac{\sigma_{LT}(\nu,Q^{2})}{Q\nu^{2}}d\nu
\end{eqnarray}
\begin{eqnarray}
\delta_{LT}(Q^{2}) = \frac{16\alpha M^{2}}{Q^{6}}\int_{0}^{x_{0}} dx~x^2\left( g_{1}+g_{2} \right).\label{eq:srdlt}
\end{eqnarray}
where $g_{LT}$ is the longitudinal-transverse interference amplitude,  $I_{LT}$ is the coefficient of the $O(\nu)$ term of the Compton amplitude, and $\sigma_{LT}$ is the longitudinal-transverse interference partial cross-section. Details on the derivations of Eqs.~\ref{eq:sr2}-\ref{eq:srdlt} can be found in~\cite{review moments}. Higher 
moments are advantageous because, thanks to their extra $x^n$ weighting, they are essentially free of the 
uncertainty associated with the low-$x$ extrapolation of the data: Reaching $x \rightarrow 0$ would 
require an infinite beam energy and hence, no data exist below a given value of $x$. Eqs.~\ref{eq:srg0} and \ref{eq:srdlt}
are examples of usage of sum rules to measure observables that are otherwise hard
to access.

In the case of the transverse-longitudinal polarizability $\delta_{LT}$, 
the $\Delta$ contribution is suppressed at low $Q^2$ because the N-$\Delta$ transition is mostly
transverse, which makes the contribution of the $\Delta$ to the longitudinal-transverse 
(LT) interference term  very small. Thus, similarly to the Bjorken sum (but with a different reason for the
$\Delta$ suppression),  $\delta_{LT}$ should also provide a robust observable to compute within the $\chi PT$ framework. Furthermore, there is the possibility that the suppression of the $\Delta$ in isovector $(p-n)$ quantities such as the Bjorken sum holds only to the first order $\Delta$ contribution $\gamma^* N \rightarrow N \pi$ but not
for the second order  $\Delta$ contribution $\gamma^* N \rightarrow \Delta \pi$. In contrast the general argument explaining the $\Delta$ suppression for  $\delta_{LT}$ should hold at all orders. If so, $\delta_{LT}$ would provide an even more robust observable than the Bjorken sum to compute in $\chi PT$. Finally, as for the Bjorken sum, the isovector part of $\gamma_{0}$, $\gamma_{0}^{p}-\gamma_{0}^{n}$, should offer similar advantages (at least for the  $\gamma^* N \rightarrow N \pi$ contribution) for checking the calculation techniques of $\chi PT$. 

\begin{figure}
\includegraphics[width=1.05\textwidth]{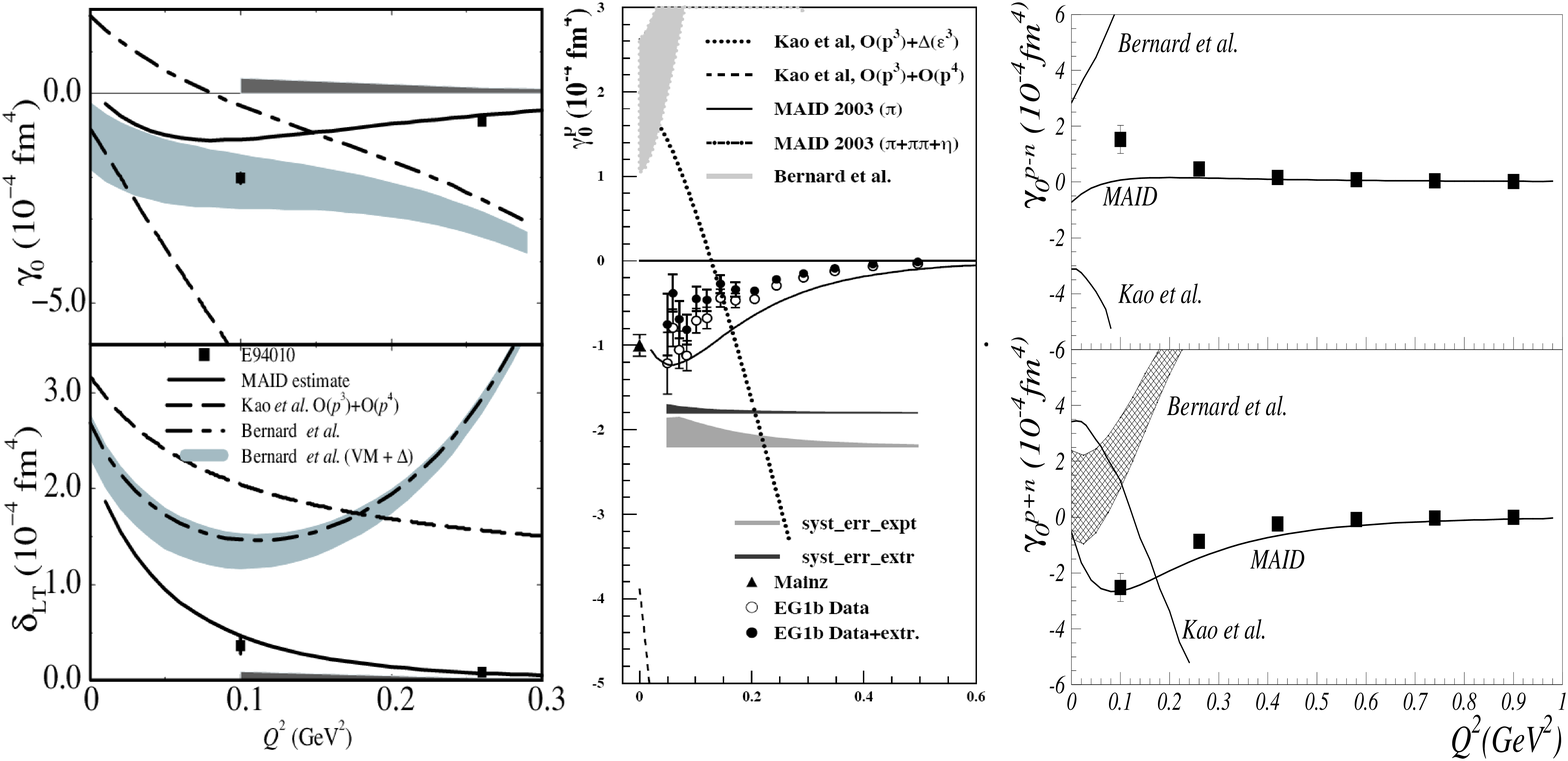}
\caption{\label{gamma0s} 
Experimental data at low $Q^2$ on generalized spin polarizabilities. Results on neutron
(JLab Hall A experiment E94010~\cite{E94010-3}) are shown on the left 
(top: $\gamma_0^n$, bottom: $\delta_{LT}^{n}$). Results on the proton (JLab CLAS 
experiment EG1b~\cite{EG1b moments}) $\gamma_0^p$ are shown in the central plot. The isospin decomposition
of $\gamma_0$ (E94010+EG1b~\cite{Bj EG1b}) is shown on the right (top: $\gamma_0^{p-n}$, bottom 
$\gamma_0^{p+n}$.}
\end{figure}

The low $Q^2$ data on forward spin polarizabilities, from Hall A E94010 and CLAS EG1b, are shown on Fig.~\ref{gamma0s}.
There is no agreement between the data and the $\chi PT$ calculations (except possibly
for the lowest $Q^2$ point of $\gamma_0^n$ that agrees with the explicitly
covariant calculation of Bernard \emph{et al}). Such disagreement is surprising 
because the lowest $Q^2$ points should be well into the validity domains of $\chi PT$.
It is even more surprising for $\gamma_0^{p-n}$ because of the $\Delta$ suppression for this quantity, and
\emph{a fortiori} for the discrepancy with $\delta_{LT}^{n}$ for which we are sure that the $\Delta$ suppression is valid at all orders . This reveals that the $\Delta$ alone is not the only cause of the lack of agreement
between data and theory, and including the $\Delta$ contribution $\gamma^* N \rightarrow \Delta \pi$ in the calculations may not be the only challenge facing $\chi PT$ calculations. In contrast, the MAID model~\cite{MAID}
is in good agreement with the data, with the notable exception of the $Q^2=0.1$ GeV$^2$ point for $\gamma_0^n$. Compared to first moments, a better agreement with higher moments 
is expected for MAID since it mostly includes one pion production reactions and these high-$x$ reactions
dominantly contribute to the higher moments.

\section{Higher moment $d_2$}
\begin{figure}
\center
\includegraphics[scale=0.15]{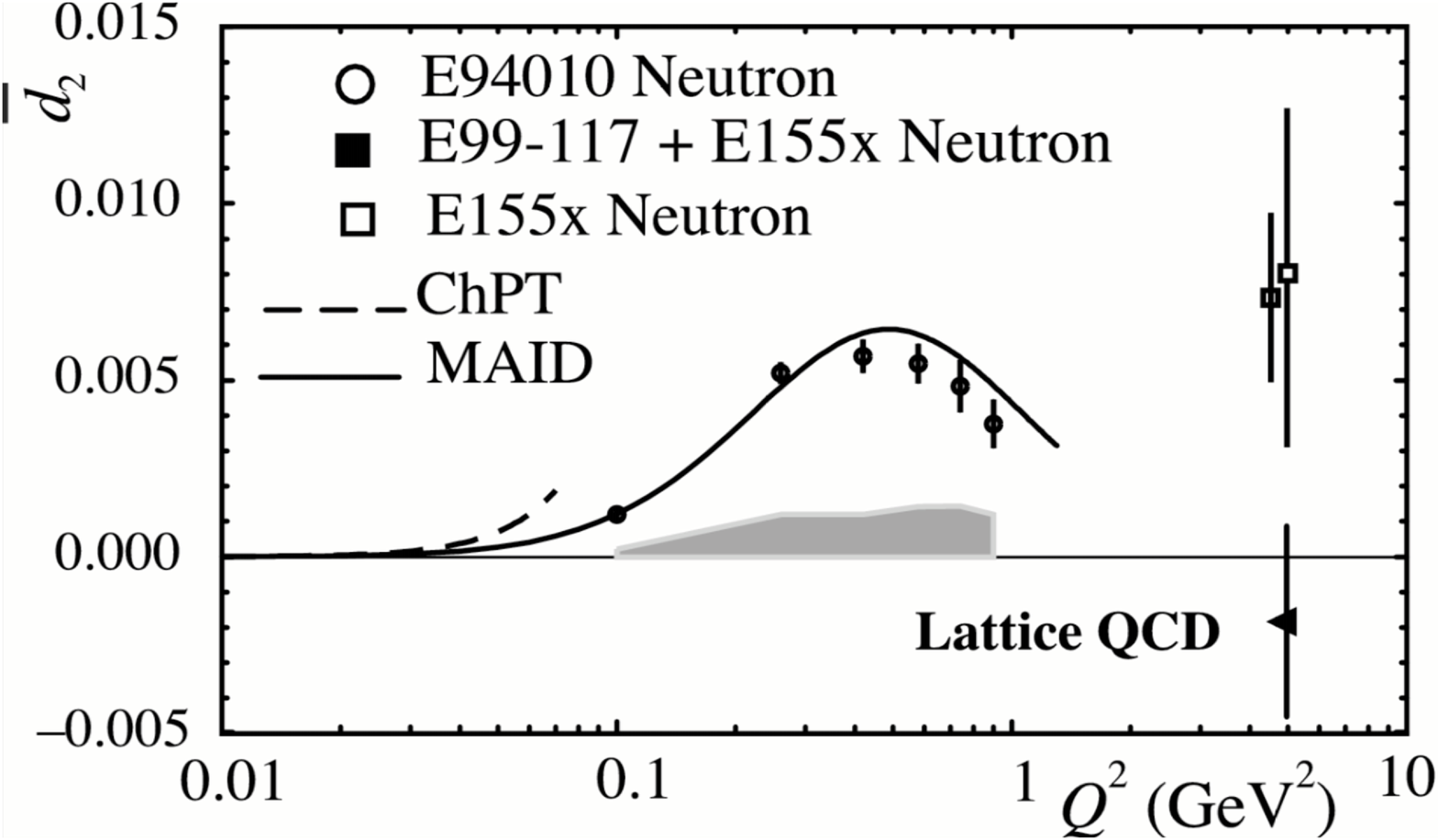}
\caption{\label{d2n} 
Results on $d_2^n$ from JLab experiments E94010, E99-117 and SLAC experiment E155x. The plain line indicates the MAID model expectation and the dashed line is for the$\chi PT$ calculations.}
\end{figure}

Another higher moment of interest is $d_2$. It can be expressed as the third moment of the twist-3 part
of $g_2$:
\begin{equation}
d_2=3 \int_{0}^{1}dx~x^2\left(g_2(x,Q^2)-g_2^{WW}(x,Q^2)\right)   \label{d2}
\end{equation} 
where $g_2^{WW}$ is the pure twist-2 part of $g_2$, first isolated by Wandzura and Wilczek~\cite{WW}.
It is a function of the leading twist expression of $g_1$:
\begin{equation}
g_2^{WW}(x,Q^2)=-g_1^{LT}(x,Q^2)+\int_{x}^{1} dy~\left(g_1^{LT}(y,Q^2)/y\right).   \label{g2ww}
\end{equation}
Hence, at large $Q^2$, $d_2$ can be cleanly interpreted as a twist-3 quantity (although, see~\cite{Accardi}). 
Its expression in term of $g_1$ and $g_2$ is:
\begin{equation}
d_2=\int_{0}^{1}dx~x^2\left(2g_1(x,Q^2)+3g_2(x,Q^2)\right)   \label{d2_2}
\end{equation} 
At intermediate $Q^2$ other higher twists contribute, while this pQCD interpretation in term of parton breaks down at low $Q^2$. We can recombine the data or the calculations on $\gamma_0$ and $\delta_{LT}$ discussed in the previous section to form $d_2$. Hence, there is no new information here regarding the data or the calculations and this recombination only recasts them in a quantity that we can cleanly interpret as a twist-3 element at large $Q^2$. 
At present, only neutron data from Hall A are available to form $d_2$ because its measurement requires a transverse target. Fig.~\ref{d2n} displays the results on $d_2^n$ from JLab experiment E94010~\cite{E94010-2}, and combined JLab E99-117~\cite{E99-117}/SLAC E155x~\cite{E155-E155x} experiments~\cite{E99-117}. 
The dashed line indicates  one of the $\chi PT$ prediction (the two calculations~\cite{meissner},\cite{Ji chipt}  yield very similar results). There is no agreement between data and the $\chi PT$ calculations. In contrast, there is again a good agreement with the MAID model expectation
(indicated by the plain line).  

\section{Summary of the $\chi PT$ checks and perspectives}
\begin{figure}
\includegraphics[scale=0.35]{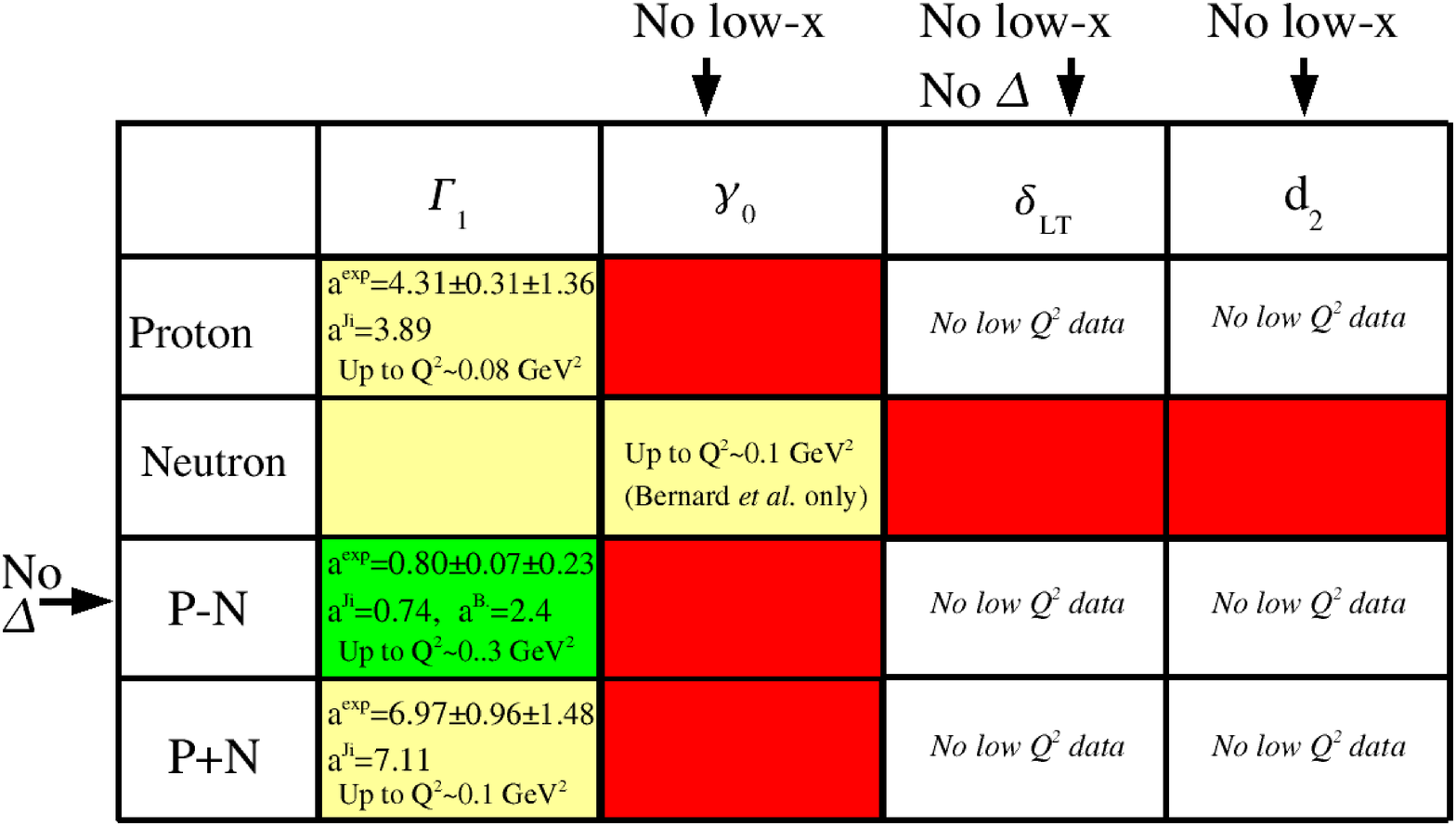}
\caption{\label{table} 
(Color online) Summary the comparison between $\chi PT$ calculations and data. The green indicates
a good match within the region in which we expect the chiral perturbation series to 
be reliable, the yellow an agreement over a shorter $Q^2$ range, and the red a mismatch. }
\end{figure}

Figure~\ref{table} summarizes the comparison between $\chi PT$ calculations and data.
It is hard to find a pattern on this summary table: It was expected that the rows or columns labelled
"no $\Delta$" or "no low-x" should have provided robust  $\chi PT$ calculations, and \emph{a fortiori} for
the intercept between the two labels ($\delta_{LT}$, $\gamma_0^{p-n}$ and $d_2^{p-n}$). Hence, the best
agreement was expected for these rows or columns. Furthermore, the fact that the $\Delta$ suppression is sure to be valid at all orders for $\delta_{LT}$ singles out this quantity as possibly the most robust calculation. Such pattern is not seen. This cannot be due to the fact that,  for first moments, the leading term is given by sum rules instead of being computed by $\chi PT$ as is done for the higher moments. This  is not the reason because for the first
moments the comparison is done on the second order term obtained from a fit to the data. The table emphasizes that more work is needed on the theoretical side (red boxes) as well as the experimental side (white boxes) for a better understanding of this problem.

The data discussed so far
were taken at JLab for experiments focused on covering the intermediate $Q^2$ 
range~\cite{eg1a proton,Bj EG1b}. A new generation of 
experiments,  E97110 in Hall A~\cite{gdh neutron}, and EG4 in Hall B~\cite{gdh proton}, 
that were especially dedicated to push such
measurements to lower $Q^2$ and higher precision, has provided new data that are 
being analyzed. During this conference, preliminary results on E97110 were presented by V. Sulkosky, while 
an update on EG4 was provided by S. Phillips. 
In addition, a new experiment to measure $\delta_{LT}^p$ in Hall A at low $Q^2$
has been approved by the JLab PAC~\cite{delta_lt_p}, while the frozen spin HD target recently arrived at 
JLab from BNL is opening new possibilities of measurements with CLAS using transversely polarized protons or deuterons.

\section{Doubly polarized pion electroproduction off proton and deuteron in the $\chi PT$ domain}

\begin{figure}
\includegraphics[scale=0.32]{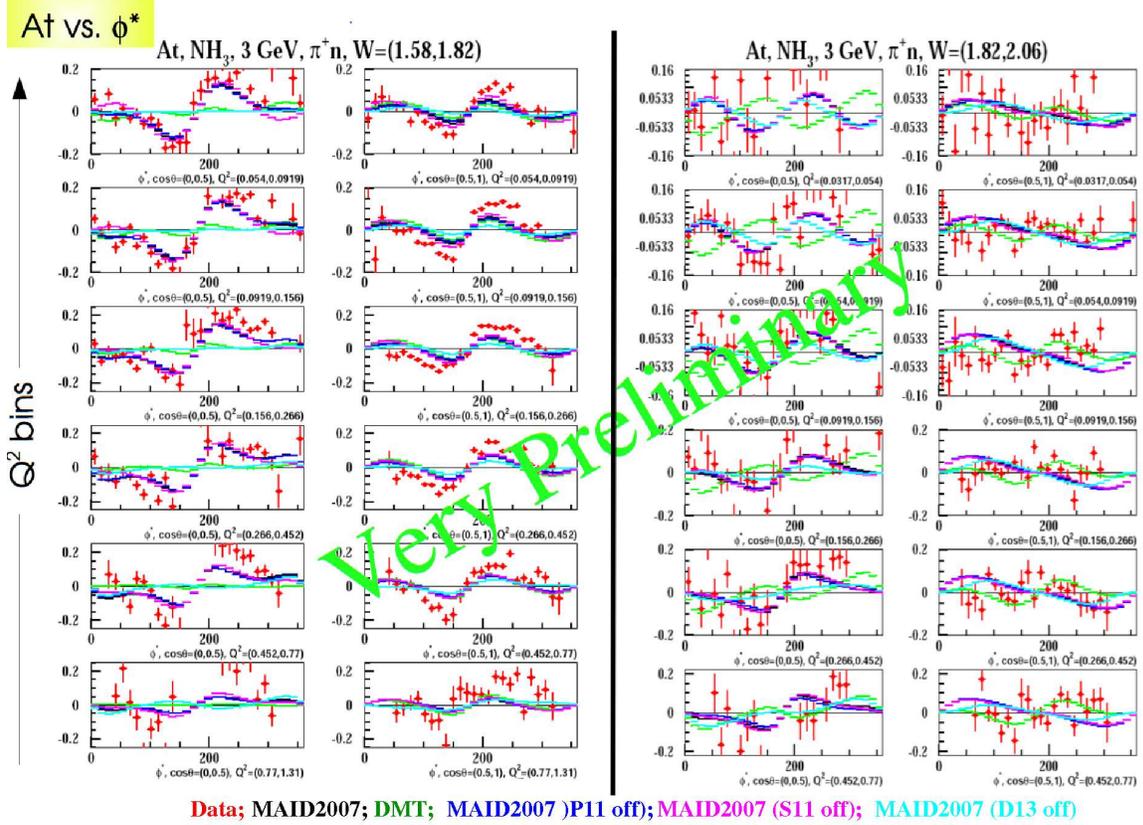}
\caption{\label{fig:pionprod} 
The preliminary $\pi^{+}$ single spin asymmetry $A_t$  from the 3 GeV  CLAS EG4 data~\cite{pionprod}.}
\end{figure}

The main goal of the CLAS EG4 experiment was to gather doubly-polarized inclusive data at low $Q^2$. However, a large quantity of (mostly single) pion electroproduction events is present. Reactions $\overrightarrow{e}\overrightarrow{p}\rightarrow e'\pi^{+}n$
and $\overrightarrow{e}\overrightarrow{p}\rightarrow e'\pi^{0}p$ are available from the NH$_3$ target and
$\overrightarrow{e}\overrightarrow{n}\rightarrow e'\pi^{-}p$ and $\overrightarrow{e}\overrightarrow{n}\rightarrow e'\pi^{0}n$ are available from the ND$_3$ target. Of the three independent asymmetries: $A_e$ (single beam asymmetry), $A_t$ (single target) asymmetry and $A_{et}$ (double beam-target asymmetry), only $A_t$ and $A_{et}$ is being analyzed since $A_e$ can be accessed more efficiently with unpolarized targets. Data are available from the four EG4 beam energies (3.0, 2.0, 1.3 and 1.1 GeV) and, by the design of the experiment, these data belong to the low $Q^2$ domain where $\chi PT$ can be applied for low enough $W$. Fig.~\ref{fig:pionprod} displays $A_t$ for  $e \overrightarrow{p}\rightarrow e'\pi^{+}n$ from the 3 GeV data in function of $\phi^*$, the angle between the scattering plane and the reaction plane. Each plot corresponds to different $Q^2$ and $W$ range.
Since the correction for the dilution from the unpolarized components of the ammonia target is not applied yet
to the data, the results from phenomenological models MAID and DMT (Dubna-Mainz-Taiwan collaboration)~\cite{DMT} are scaled down by
a factor 0.2, the approximately expected dilution of the data. No acceptance corrections are applied yet in this analysis and those might modify the features seen on Fig.~\ref{fig:pionprod}.  Nevertheless, the models and preliminary data are qualitatively similar. The important information from these plots, that presents only a 
fraction of the analyzed data, is
the quality of the data accuracy, the extensive kinematic range on which data are available, and the fact
that part of the data are taken in the applicability domain of $\chi PT$. No $\chi PT$ calculations are available yet
for such reactions. The wealth and quality of data warrant thorough tests of $\chi PT$ if such calculations
become available. 

\section{Summary and perspectives}
We discussed the Jefferson Lab data on moments of spin structure functions at large distances
and compared them to $\chi PT$, the effective theory of strong force that should describe it at large 
distances. The data and calculations do not consistently agree. In particular, the better agreement expected for observables in which the $\Delta$ resonance is suppressed is seen only for the Bjorken sum, but not for $\gamma_{0}^{p}-\gamma_{0}^{n}$ nor for $\delta_{LT}^n$ even if for this latest quantity we are sure that the $\Delta$ is suppressed at all orders, while this is not certain for the isovector $(p-n)$ quantities. Apparently, the $\Delta$ cannot explain single-handedly the discrepancy between data and calculations. The new generation of experiments that gathered data at lower $Q^2$,  E97110 and EG4  might help shed light on this problem.  In addition, data from transversely polarized targets are likely to be crucial to solve the puzzle. These data should be provided by the  new E08027 experiment to measure $\delta_{LT}^p$ in Hall A at low $Q^2$, and possibly by electron scattering experiments using the Hall B frozen spin HD target with CLAS. The analysis of the  large amount of doubly polarized pion electroproduction data in the low $Q^2$ domain from the EG4 experiment is  well advanced.  The preliminary results are being compared to phenomenological models. $\chi PT$ predictions are not available so far for these observables but  would be very valuable.

\end{document}